\documentstyle[sprocl,epsf,pstricks]{article}

\def\NPB{{\em Nucl. Phys.} B} 
\def\PLB{{\em Phys. Lett.}  B} 

\def\PRD{{\em Phys. Rev.} D} 
 
\def\be{\begin{equation}} 
\def\ee{\end{equation}} 
\def\bea{\begin{eqnarray}} 
\def\eea{\end{eqnarray}} 

\newcommand{\laem}{\begin{array}{c} < \vspace{-.85em} \\ {\scriptstyle \sim}
\end{array}}
\newcommand{\gaem}{\begin{array}{c} > \vspace{-.85em} \\ {\scriptstyle \sim}
\end{array}}
\newcommand{\beq}{\begin{equation}}
\newcommand{\eeq}{\end{equation}}
\newcommand{\beqa}{\begin{eqnarray}}
\newcommand{\eeqa}{\end{eqnarray}}

\def\slashchar#1{\setbox0=\hbox{$#1$}           
   \dimen0=\wd0                                 
   \setbox1=\hbox{/} \dimen1=\wd1               
   \ifdim\dimen0>\dimen1                        
      \rlap{\hbox to \dimen0{\hfil/\hfil}}      
      #1                                        
   \else                                        
      \rlap{\hbox to \dimen1{\hfil$#1$\hfil}}   
      /                                         
   \fi}                                         %

\newcmykcolor{purp}{.2 1 0 .4}
\newcmykcolor{grnn}{1 .2 .5 .4}

\newcommand{\pr}{\em Phys.\ Rev.\ }
\newcommand{\prp}{\em Phys.\ Rep.\ }

\newcommand{\pl}{\em Phys.\ Lett.\ {\bf B}}

\begin{document} 
 
\title{Custodial Symmetry, Flavor Physics, and the \\
Triviality Bound on the Higgs Mass} 
 
\author{R.~S.~Chivukula, E.~H.~Simmons, \& B.~A.~Dobrescu} 
 
\address{Department of Physics, Boston University\\ 
590 Commonwealth Ave., Boston MA 02215 USA \\
{\tt sekhar@bu.edu, simmons@bu.edu, dobrescu@budoe.bu.edu}\\
{\tt BUHEP-97-7 and hep-ph/9703206}}
 
 
\maketitle\abstracts{ The triviality of the scalar sector of the
  standard one-doublet Higgs model implies that this model is only an
  effective low-energy theory valid below some cut-off scale $\Lambda$.
  We show that the experimental constraint on the amount of custodial
  symmetry violation implies that the scale $\Lambda$ must be greater
  than of order 7.5 TeV.  The underlying high-energy theory must also
  include flavor dynamics at a scale of order $\Lambda$ or greater in
  order to give rise to the different Yukawa couplings of the Higgs to
  ordinary fermions.  This flavor dynamics will generically produce
  flavor-changing neutral currents. We show that the experimental
  constraints on the neutral $D$-meson mass difference imply that
  $\Lambda$ must be greater than of order 21 TeV.  For theories defined
  about the infrared-stable Gaussian fixed-point, we estimate that this
  lower bound on $\Lambda$ yields an upper bound of approximately 460
  GeV on the Higgs boson's mass, independent of the regulator chosen to
  define the theory. We also show that some regulator schemes, such as
  higher-derivative regulators, used to define the theory about a
  different fixed-point are particularly dangerous because an infinite
  number of custodial-isospin-violating operators become relevant.}

\section{Triviality and the Standard Model\,\protect{\cite{rscintro}}}

In the standard Higgs model, one introduces
a fundamental scalar doublet:
\beq
\phi=\left(\matrix{\phi^+ \cr \phi^0 \cr}\right)
{}~,
\eeq
with potential:
\beq
V(\phi)=\lambda \left(\phi^{\dagger}\phi - {v^2\over 2}\right)^2
{}~.
\label{eq:pot}
\eeq
While this theory is simple and renormalizable, it has
a number of shortcomings. First, while the theory
can be constructed to accommodate the
breaking of electroweak symmetry, it provides no {\it explanation} 
for it -- one simply assumes that the potential
is of the form in eqn.~(\ref{eq:pot}). In addition, in
the absence of supersymmetry, quantum corrections to
the Higgs mass are naturally of order the largest scale
in the theory
\beq
{\lower5pt\hbox{\epsfysize=0.25 truein \epsfbox{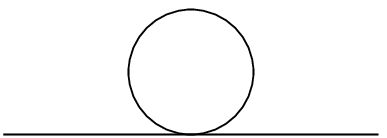}}}
\Rightarrow  m_H^2 \propto \Lambda^2~,
\eeq
leading to the hierarchy and naturalness problems.\cite{thooft} 
Finally, the $\beta$ function for the self-coupling $\lambda$
\beq
{\lower7pt\hbox{\epsfysize=0.25 truein \epsfbox{figures/beta.eps}}}
\Rightarrow \beta = {3\lambda^2 \over 2 \pi^2} \, > \, 0
{}~,
\eeq
leading to a ``Landau pole'' and triviality.\cite{trivial}

\begin{figure}[tbp]
\centering
\epsfysize=2in
\hspace*{0in}
\epsffile{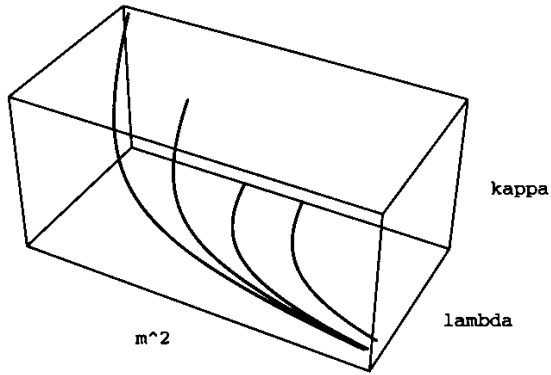}
\caption{Renormalization group flow of Higgs mass $m^2$, Higgs
self-coupling $\lambda$, and the coefficient of a representative
irrelevant operator $\hat{\kappa}$. The flows go from upper-left to
lower-right as one scales to the infrared.}
\label{Fig1}
\end{figure}

The hierarchy/naturalness and triviality problems
can be nicely summarized in terms of the 
Wilson renormalization group.
Define the theory with a fixed UV-cutoff:
\beqa
{\cal L}_\Lambda =  & D^\mu \phi^\dagger D_\mu \phi + 
m^2(\Lambda)\phi^\dagger \phi 
+ {\lambda(\Lambda)\over 4}(\phi^\dagger\phi)^2 \nonumber\\
& + {\hat{\kappa}(\Lambda)\over 36\Lambda^2}(\phi^\dagger\phi)^3+\ldots  
\label{eq:liz}
\eeqa
Here $\hat{\kappa}$ is the coefficient of a representative 
irrelevant operator, 
of dimension greater than four.
Next, integrate out states with $\Lambda^\prime < k < \Lambda$,
and construct a new Lagrangian with the same {\it
low-energy} Green's functions:
\beqa
{\cal L}_\Lambda & \Rightarrow & {\cal L}_{\Lambda^\prime} \nonumber\\
m^2(\Lambda)& \rightarrow & m^2(\Lambda^\prime) \nonumber \\
\lambda(\Lambda) & \rightarrow & \lambda(\Lambda^\prime) \nonumber \\
\hat{\kappa}(\Lambda) & \rightarrow & \hat{\kappa}(\Lambda^\prime)  
\eeqa
The low-energy behavior of the theory is then nicely summarized in terms
of the evolution of couplings in the infrared.\footnote{For convenience,
  we ignore the corrections due to the weak gauge interactions.  In
  perturbation theory, at least, the presence of these interactions does
  not qualitatively change the features of the Higgs sector.} A
three-dimensional representation of this flow in the
infinite-dimensional space of couplings shown in Figure \ref{Fig1}.

From Figure \ref{Fig1}, we see that as we scale to the infrared the
coefficients of irrelevant operators, such as $\hat{\kappa}$, tend to
zero; {\it i.e.} the flows are attracted to the finite dimensional
subspace spanned (in perturbation theory) by operators of dimension four
or less; this is the modern understanding of {\it renormalizability}. On
the other hand, the coefficient of the only {\it relevant} operator (of
dimension 2), $m^2$, tends to infinity. This leads to the
naturalness/hierarchy problem.\cite{thooft} Since we want $m^2 \propto
v^2$ at low energies we must adjust the value of $m^2(\Lambda)$ to a
precision of
\beq
{\Delta m^2(\Lambda) \over m^2(\Lambda)} \propto {v^2 \over \Lambda^2}~.
\eeq

Central to our discussion here is
the fact that the coefficient of the only
marginal operator $\lambda$ tends to 0, because of the positive $\beta$
function.  If we try to take the continuum limit, $\Lambda \to
+\infty$, the theory becomes free or trivial.\cite{trivial} The
triviality of the scalar sector of the standard one-doublet Higgs model
implies that this theory is only an effective low-energy theory valid
below some cut-off scale $\Lambda$.  Physically this scale marks the
appearance of new strongly-interacting symmetry-breaking dynamics. 
Examples of such high-energy theories include ``top-mode''
standard models\,\cite{topmode} and composite Higgs models.\cite{chiggs}
As the Higgs mass increases, the upper bound on the scale $\Lambda$
decreases.  An estimate of this effect can be obtained by
integrating the one-loop $\beta$-function, which yields
\beq
\lambda(m_H) \stackrel{<}{\sim} {{2\pi^2}\over 3\log{\Lambda\over m_H}}\, .
\label{eq:est}
\eeq

Using the relation $m^2_H = 2\lambda(m_H) v^2$ we find 
\beq 
m^2_H \ln\left({\Lambda\over m_H}\right)\le {4\pi^2 v^2 \over 3}~.
\label{estimate}
\eeq 
Hence a lower bound\,\cite{cabbibo,dashen} on $\Lambda$ yields an upper
bound on $m_H$. We must require that $M_H / \Lambda$ in
eqn.~(\ref{estimate}) be small enough to afford the effective Higgs
theory some range of validity (or to minimize the effects of
regularization in the context of a calculation in the scalar theory).
Quantitative\,\cite{lattice} studies on the lattice using analytic and
Monte Carlo techniques result in an upper bound on the Higgs mass of
approximately 700 GeV.  The lattice Higgs mass bound is potentially
ambiguous because the precise value of the bound on the Higgs boson's
mass depends on the (arbitrary) restriction placed on $M_H / \Lambda$.
The ``cut-off'' effects arising from the regulator are not universal:
different schemes can give rise to different effects of varying sizes
and can change the resulting Higgs mass bound.

In this talk we show that, for models that reproduce the standard
one-doublet Higgs model at low energies, electroweak and flavor
phenomenology provide a lower bound on the scale $\Lambda$ of order 10
-- 20 TeV that is regularization-independent (i.e.  independent of the
details of the underlying physics). Using eqn.~(\ref{estimate}) we
estimate that this gives an {\it upper} bound of 450 -- 500 GeV on the
Higgs boson mass.

The discussion we have presented is based on perturbation theory and is
valid in the domain of attraction of the ``Gaussian fixed point''
($\lambda=0$).  In principle, however, the Wilson approach can be used
{\it non-perturbatively}, even in the presence of
nontrivial fixed points or large anomalous dimensions.  In a
conventional Higgs theory, neither of these effects is thought to
occur.\cite{lattice} We return to the issue of the possible existence
of other, potentially non-trivial, fixed points in section 4 below.

\section{Dimensional Analysis}

We will analyze the effects of the underlying physics by estimating the
sizes of various operators in a low-energy effective lagrangian
containing the (presumably composite) Higgs boson and the ordinary gauge
bosons and fermions. Since we are considering theories with a heavy
Higgs field, we expect that the underlying high-energy theory will be
strongly interacting. Borrowing a technique from QCD we will rely on
dimensional analysis\,\cite{QCDNDA} to estimate the sizes of various
effects of the underlying physics.

A strongly interacting theory has no small parameters.  As noted by
Georgi,\cite{generalized} a theory\,\footnote{These dimensional estimates
  only apply if the low-energy theory, when viewed as a scalar field
  theory, is defined about the infrared-stable Gaussian fixed-point. We
  return to potentially ``non-trivial'' theories below.} with light
scalar particles belonging to a single symmetry-group representation
depends on two parameters: $\Lambda$, the scale of the underlying
physics, and $f$ (the analog of $f_\pi$ in QCD), which measures the
amplitude for producing the scalar particles from the vacuum. Our
estimates will depend on the ratio $\kappa = \Lambda / f$, which is
expected to fall between 1 and $4\pi$.

Consider the kinetic energy of a scalar bound-state in the appropriate
low-energy effective lagrangian. The properly normalized kinetic energy
is
\beq
\partial^\mu \phi^\dagger \partial_\mu \phi
= { \Lambda^2 f^2} \left({\partial^\mu \over {  \Lambda}}\right)
\left({\phi^\dagger \over {  f}}\right)
\left({\partial_\mu \over {  \Lambda}}\right) 
\left({\phi \over {  f}}\right)~,
\eeq
where, because the fundamental scale of the interactions is $\Lambda$,
we ascribe a $\Lambda$ to each derivative and an $f$ to each $\phi$
since $f$ measures the amplitude to produce the bound state. This tells
us that the overall magnitude of each term in the effective lagrangian is
${\cal{O}}(f^2\Lambda^2)$.  We can next estimate the  ``generic'' size
of a mass term in the effective theory:
\beq
m^2 \phi^\dagger \phi = { \Lambda^2 f^2} \left({\phi^\dagger 
\over { f}}\right)
\left({\phi \over { f}}\right) \Rightarrow 
{m^2 \propto \Lambda^2}~.
\eeq
This is a reproduction of the hierarchy problem.  In the absence of some
other symmetry not accounted for in these rules, fine-tuning\,\footnote{We
  will not be addressing the hierarchy problem here; we will simply
  assume that some other symmetry or dynamics has produced the
  appropriate light scalar state.} is required to obtain $m^2 \ll
\Lambda^2$. Next, consider the size of scalar interactions.  From the
simplest interaction
\beq
\lambda (\phi^\dagger \phi)^2 \Rightarrow {  \lambda 
\propto \left({\Lambda \over f}\right)^2 = \kappa^2}~,
\eeq
we see that $\kappa$ will determine the size of coupling
constants. Similarly, for a higher-dimension interaction such as
the one in eqn.~(\ref{eq:liz}) we find
\beq
{\hat{\kappa} \over {  \Lambda^2}}(\phi^\dagger \phi)^3 \Rightarrow
{  \hat{\kappa} \propto \kappa^4}~.
\eeq

These rules are easily extended to include strongly-interacting fermions
self-consistently.  Again, we start with the properly normalized
kinetic-energy
\beq
\bar{\psi}\slashchar{\partial}\psi = 
{ \Lambda^2 f^2} \left({\bar{\psi} \over { f\sqrt{\Lambda}}}\right)
\left({\slashchar{\partial} \over { \Lambda}}\right)
\left({\psi \over { f\sqrt{\Lambda}}}\right)~,
\eeq
and learn that $f\sqrt{\Lambda}$ is a measure of the amplitude
for producing a fermion from the vacuum. Next, consider
a Yukawa coupling of a strongly-interacting fermion to
our composite Higgs,
\beq
y (\bar{\psi}\phi \psi) \Rightarrow {  y \propto \kappa}~.
\label{eq:natyukawa}
\eeq
And finally, the natural size of a four-fermion operator is
\beq
{\nu \over {  \Lambda^2}} (\bar{\psi}\psi)^2 \Rightarrow 
{  \nu \propto \kappa^2}~.
\label{eq:grhoex}
\eeq

We will rely on these estimates to derive bounds on the scale $\Lambda$.
By way of justification, we note that these estimates work in QCD for the
chiral-Lagrangian,\cite{QCDNDA} with $f \to f_\pi$, $\Lambda \to 1$ GeV,
and $\kappa \approx {\cal{O}}(4 \pi)$.  For example, four nucleons
operators of the form shown in eqn.~(\ref{eq:grhoex}) arise in the vector
channel from $\rho$-exchange and we obtain $\Lambda = m_\rho$ and
$\kappa = g_\rho \approx 6$.  In a QCD-like theory with $N_c$ colors and
$N_f$ flavors one expects\,\cite{reconsider} that
\beq
\kappa \approx \min \left({4\pi a\over N_c^{1/2}},
{4\pi b\over N_f^{1/2}}\right)~,
\eeq
where $a$ and $b$ are constants of order 1. In the results that
follow, we will display the dependence on $\kappa$ explicitly; when
giving numerical examples, we set $\kappa$ equal to the geometric mean
of 1 and $4\pi$, {\it i.e.} $\kappa \approx 3.5$.

\section{Isospin Violation and Bounds\,\protect{\cite{rscehs}} on $m_H$}

Because of the $SU(2)_W \times U(1)_Y$ symmetry of the low-energy
theory, all terms of dimension less than or equal to four respect
custodial symmetry.\cite{custodial} The leading custodial-symmetry
violating operator is of dimension six\,\cite{wyler,grinstein} 
and involves four Higgs doublet fields
$\phi$. According to the rules of dimensional analysis, the operator
\beq
{\lower35pt\hbox{\epsfysize=1.0 truein \epsfbox{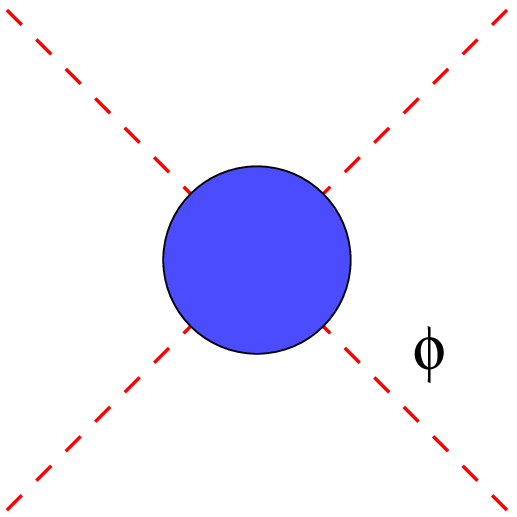}}}
\Rightarrow
{\kappa^2 \over \Lambda^2} 
(\phi^\dagger D^\mu \phi)
(\phi^\dagger D_\mu \phi)~,
\label{eq:isoviol}
\eeq
should appear in the low-energy effective theory with a coefficient
of order one.\cite{grinstein} Such an operator will give rise to a deviation
\beq
\Delta \rho_* = - {\cal O}\left(\kappa^2 {v^2 \over \Lambda^2}\right) ~,
\eeq
where $v \approx 246$ GeV is the expectation value of the Higgs
field. Imposing the constraint\,\cite{data,something} 
that $|\Delta \rho_*| \le 0.4\%$, we find the lower bound
\beq
\Lambda \stackrel{>}{\sim} 4\, {\rm TeV} \cdot \kappa ~.
\eeq
For $\kappa \approx 3.5$, we find $\Lambda \stackrel{>}{\sim} 14$ TeV. 

Alternatively, it is possible that the underlying strongly-interacting
dynamics respects custodial symmetry. Even in this case, however, there
must be custodial-isospin-violating physics (analogous to
extended-technicolor\,\cite{Lane,Dimopoulos} interactions) which couples the
$\psi_L=(t,\ b)_L$ doublet and $t_R$ to the strongly-interacting
``preon'' constituents of the Higgs doublet in order to produce a top
quark Yukawa coupling at low energies and generate the top quark mass.
If, for simplicity, we assume that these new weakly-coupled
custodial-isospin-violating interactions are gauge interactions with
coupling $g$ and mass $M$, dimensional analysis allows us to estimate
the size of the resulting top quark Yukawa coupling.  The ``natural
size'' of a Yukawa coupling (eqn.~(\ref{eq:natyukawa})) is $\kappa$ and
that of a four-fermion operator (eqn.~(\ref{eq:grhoex})) is
$\kappa^2/\Lambda^2$; the ratio $(g^2/M^2)/(\kappa^2/\Lambda^2)$ is the
``small parameter'' associated with the extra flavor interactions and we
find
\beq {\lower35pt\hbox{\epsfysize=1.0 truein
    \epsfbox{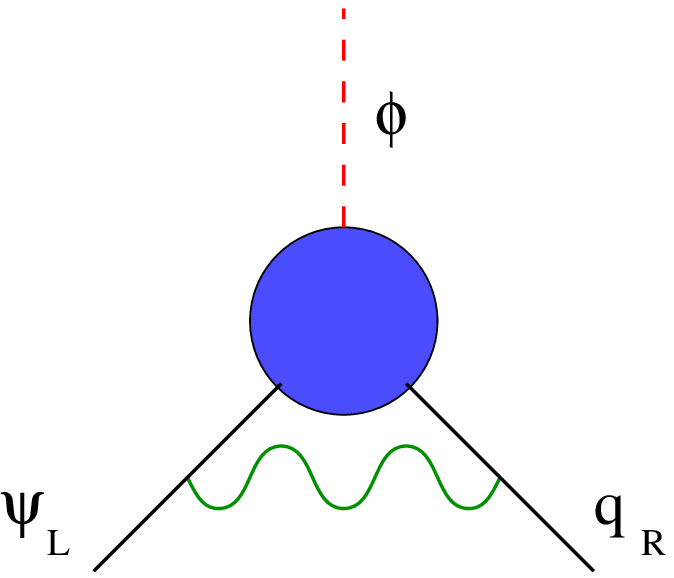}}} \Rightarrow {g^2 \over M^2} {\Lambda^2 \over
  \kappa}\bar{q}_R \phi \psi_L ~.
\label{eq:quarkpreon}
\eeq
In order to give rise to a quark mass $m_q$, the 
Yukawa coupling must be equal to
\beq
{\sqrt{2} m_q \over v}
\eeq
where $v\approx 246$ GeV. This implies
\beq
\Lambda \stackrel{>}{\sim}{M \over g} \sqrt{\sqrt{2} \kappa {m_q \over v}}~.
\label{eq:yukawa}
\eeq 

These new gauge interactions will typically also give rise to
custodial-isospin-violating 4-preon interactions\,\footnote{These
  interactions have previously been considered in the context of
  technicolor theories.\cite{appelquist}} which, at low energies, will
give rise to an operator of the same form as the one in
eqn.~(\ref{eq:isoviol}). Using dimensional analysis, we find
\beq
{\lower35pt\hbox{\epsfysize=1.0 truein \epsfbox{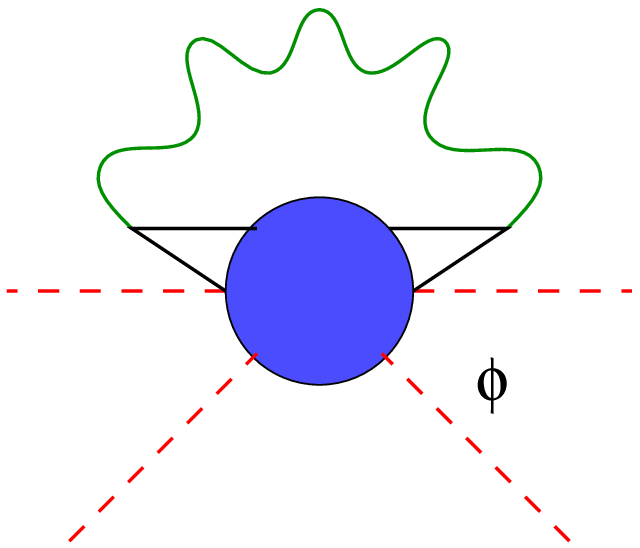}}}
\Rightarrow 
\left[{  {g^2 \over M^2}}
\left({  {\kappa^2 \over \Lambda^2}}\right)^{-1}\right] 
{  \kappa^2 \over \Lambda^2} (\phi^\dagger D^\mu \phi) (\phi^\dagger
D_\mu \phi)~,
\label{eq:isoviola}
\eeq
which results in the bound  $M/g \stackrel{>}{\sim} 4$ TeV.
From eqn.~(\ref{eq:yukawa}) with $m_t \approx 175$ GeV we then derive the limit
\beq
\Lambda \stackrel{>}{\sim} 4\, {\rm TeV} \cdot \sqrt{\kappa}~.
\eeq
For $\kappa \approx 3.5$, we find $\Lambda \stackrel{>}{\sim} 7.5$ TeV.

\section{Non-Trivial Scaling}

Dimensional analysis was crucial to the discussion given above.  If the
low-energy Higgs theory does not flow toward the trivial Gaussian
fixed-point in the infrared limit, the scaling dimensions of the fields
and operators can be very different than naively expected.  In this case
the bounds given above do not apply.

A nice example of a scalar theory with non-trivial behavior has been
given by Jansen, Kuti, and Liu.\cite{jansen} They
consider a theory defined by an  $O(4)$-symmetric Lagrange density
with a modified kinetic-energy
\beq
{\cal L}_{kin} = -{1\over 2} \phi^\dagger (\Box + {\Box^3\over {\cal M}^4})
\phi~.
\eeq
In the large-$N$ limit, this higher-derivative kinetic term is
sufficient to eliminate all divergences. A lattice simulation of this
theory\,\cite{liu} indicates that this approach can be used to define a
non-trivial Higgs theory with a Higgs boson mass as high as 2 TeV, while
avoiding any noticeable effects from the (complex-conjugate) pair of
ghosts which are present because of the higher derivative kinetic-energy
term. 

As shown by Kuti,\cite{kutii} in the infrared this higher-derivative
theory flows to a non-trivial fixed point on an infinite dimensional
critical surface, which corresponds to a continuum field theory with an
infinite number of relevant operators. The reason there are an infinite
number of relevant operators is that, if the continuum limit is taken so
that the scale ${\cal M}$ remains finite as required in order to flow
to a non-trivial theory, the scaling dimension\,\cite{kutii} of the
Higgs doublet field $\phi$ is -1 instead of the canonical value of
+1! 

If one could impose an exact $O(4)$ symmetry on the symmetry breaking
sector, this would lead to a strongly-interacting electroweak
symmetry-breaking sector without technicolor\,\cite{liu}.  However, as
argued above, custodial isospin violation in the flavor sector must
couple to the 
symmetry-breaking sector to give rise to the different top- and
bottom-quark masses. Furthermore, if the scaling dimension of the
Higgs field is -1, there is an infinite class of custodial-isospin-violating
operators (including the operator in eqn.~(\ref{eq:isoviol})) which are
relevant. Since these operators are relevant, even a small amount of 
custodial isospin violation coming from high-energy flavor dynamics will
be amplified as one scales to low energies, ultimately contradicting
the bound on $\Delta\rho_*$. We therefore conclude
that these non-trivial scalar theories cannot provide a phenomenologically 
viable theory of electroweak symmetry breaking. 

To construct a phenomenologically viable theory of a
strongly-interacting Higgs sector it is not sufficient to
construct a theory with a heavy Higgs boson; one must also ensure that
all potentially custodial-isospin-violating operators remain
irrelevant.\footnote{This is also a concern in walking
  technicolor.\cite{rsc}}

\section{Flavor-Changing Neutral-Currents\,\protect{\cite{rscbadehs}}}

The high-energy flavor physics responsible for the generation of the
quark-preon couplings {\it must} distinguish between different flavors
so as to give rise to the different masses of the corresponding
fermions.  In addition to the Higgs-fermion coupling discussed above,
the flavor physics will also give rise to flavor-specific couplings of
ordinary fermions to themselves.  These new current-current interactions
among ordinary fermions generically give rise to flavor-changing neutral
currents (as previously noted\,\cite{Lane} for the case of ETC theories)
that affect Kaon and $D$-meson physics.  For instance, consider the
interactions responsible for the $s$-quark mass. Through Cabibbo mixing,
these interactions must couple to the $d$-quark as well.  This will
give rise to the interactions
\beqa
{\cal L}_{eff} & = & - \, (\cos \theta_L^s \sin \theta_L^s)^2 
\frac{g^{2}}{M^{2}}
( \overline s_L  \gamma^{\mu} d_L )(\overline s_L  \gamma_{\mu} d_L)
\nonumber \\ [2mm]
& & -  \, (\cos \theta_R^s \sin \theta_R^s)^2 
\frac{g^{2}}{M^{2}}
( \overline s_R \gamma^{\mu} d_R)(\overline s_R \gamma_{\mu} d_R)
\nonumber \\ [2mm]
& & - \,
\cos \theta_L^s \sin \theta_L^s \cos \theta_R^s \sin \theta_R^s
\frac{g^{2}}{M^{2}}
( \overline s_L  \gamma^{\mu} d_L )(\overline s_R \gamma_{\mu} d_R)~,
\label{ops1}
\eeqa
where the coupling $g$ and mass $M$ are of the same order as those
in the interactions which ultimately give rise to the $s$-quark
Yukawa coupling in eqn.~(\ref{eq:quarkpreon}), and the angles
$\theta^s_L$ and $\theta^s_R$ represent the relation between the
gauge eigenstates and the mass eigenstates.  The operators in
eqn.~(\ref{ops1}) will clearly affect neutral Kaon physics.  Similarly, the
interactions responsible for other quarks' masses will give rise to
operators that contribute to mixing and decays of various mesons.

Since the operators responsible for generating quark masses and for
causing flavor-changing neutral currents violate flavor symmetries
differently,\cite{ctsm} in principle one could construct a theory with
an approximate GIM symmetry.\cite{ctsm,oldtcgim,technigim} In such
models, flavor-changing neutral currents would be suppressed but
different quarks would still receive different masses.  A theory of this
type which included a light scalar state (unlike previous
examples\,\cite{ctsm,oldtcgim,technigim}) would be able to evade the
flavor-changing neutral current limits discussed here.

\subsection{Flavor-Changing Neutral Currents: $\Delta S$}

To start, let us consider the
four-fermion interactions in eqn.~(\ref{ops1}), which will alter the
predicted value of the $K_L - K_S$ mass difference.  Using the
vacuum-insertion approximation,\cite{vacinsert} we can estimate
separately how much the purely left-handed (LL), purely right-handed
(RR) and mixed (LR) current-current operators contribute.  Requiring
each contribution to be less than the observed mass difference $\Delta
m_K$, we find the bounds
\beqa
\left(\frac{M}{g}\right)_{\! {\rm LL,RR}} \! & \gaem &\!
f_K \left( \frac{2  m_K B_K}{3 \Delta m_K }\right)^{\! 1/2}
\cos \theta_{L,R}^s \sin \theta_{L,R}^s
\\ [2mm]
& \approx & \! 0.92 \times 10^{3} \, {\rm TeV} 
\cos \theta_{L,R}^s \sin \theta_{L,R}^s
\eeqa
from the first two operators in eqn.~(\ref{ops1}), and
\beqa
\hspace*{-8mm}
\!\!\left(\frac{M}{g}\right)_{\! {\rm LR}}\!\!
&\!\gaem\! & \!
f_K \left\{ \frac{m_K B_K^{\prime}}{3 \Delta m_K } 
\left[ \frac{m_K^2}{(m_s + m_d)^2} - \frac{3}{2} \right] 
\right\}^{\! 1/2}\!\!\!
(\cos \theta_L^s \sin \theta_L^s \cos \theta_R^s \sin \theta_R^s)^{\! 1/2}
\nonumber\\ [2mm]
& \! \approx \! & \! 1.4 \times 10^{3} \, {\rm TeV} \,
(\cos \theta_L^s \sin \theta_L^s \cos \theta_R^s \sin \theta_R^s)^{\! 1/2}
\eeqa
from the last operator in eqn.~(\ref{ops1}). In evaluating these
expressions, we have used the values $f_K \approx 113$ MeV, the ``bag''
factors $B_K, B_K^\prime \sim 0.7$, and $m_s + m_d \sim 200$ MeV.  In
order to produce the observed $d - s$ mixing, we expect that at least one of the
angles $\theta_L^s,\ \theta_R^s$ is of order the Cabibbo angle,
$\theta_C$.  Then we find from any one operator that
\beq
\frac{M}{g}
\gaem 200 \, {\rm TeV}~.
\label{fcncmbound}
\eeq
From eqn.~(\ref{eq:yukawa}) it follows that
\beq
\Lambda \gaem 6.8 \, {\rm TeV} 
\sqrt{\kappa\left({m_s\over 200\, {\rm MeV}}\right)}~.
\label{eq:fcncbound}
\eeq
For $\kappa\approx 3.5$, this yields a
lower bound of approximately 13 TeV on $\Lambda$.

Typically, in addition to the operators in eqn.~(\ref{ops1}) there will
be flavor-changing operators which are products of color-octet
currents\,\footnote{ Note that it is likely that color must be embedded in
  the flavor interactions in order to avoid possible Goldstone
  bosons\,\protect\cite{Lane} and large contributions to the $S$
  parameter.\protect\cite{Spar}}.  At least in the vacuum-insertion
approximation, the matrix elements of products of color-octet currents
are enhanced relative to those shown in (\ref{ops1}) by a factor of 4/3
for the LL and RR operators and a factor of approximately 7 for the LR
operator.  Furthermore, because left-handed quarks are weak doublets it
is possible that flavor physics associated with the $c$-quark mass also
contributes to $\Delta S = 2$ interactions. If so, one would replace
$m_s$ with $m_c$ in eqn.~(\ref{eq:fcncbound}), yielding a lower bound on
$\Lambda$ of order 20$\sqrt{\kappa}$ TeV.  For these reasons, the bounds
given above may be conservative.

\subsection{Flavor-Changing Neutral Currents: $\Delta C$}

Usually, the strongest constraints on nonstandard physics from
flavor-changing neutral currents come from processes involving Kaons,
like those considered above.  In the present case, however, the
constraints from $D^0 - \overline{D}^0$ mixing are also important
because the $c$-quark is heavier than the $s$-quark, while the $u-c$
mixing is as large as the $d-s$ mixing.

Again, there are contributions to $D$-meson mixing from the
color-singlet products of currents analogous to those in
eqn.~(\ref{ops1}). The purely left-handed or right-handed
current-current operators yield
\beq
\left(\frac{M}{g}\right)_{ \! {\rm LL,RR}} 
\gaem 
f_D\left( \frac{2  m_D B_D}{3 \Delta m_D }\right)^{\! 1/2}
\cos \theta_{L,R}^c \sin \theta_{L,R}^c \approx 120 \, {\rm TeV} ~,
\eeq
where we have used the limit\,\cite{data} on the neutral $D$-meson mass
difference, $\Delta m_D \laem 1.4 \times 10^{-10}$ MeV, and $f_D
\sqrt{B_D} = 0.2$ GeV, $\theta_{L,R}^c \approx \theta_C$.  The bound on
the scale of the underlying strongly-interacting dynamics follows from
eqn.~(\ref{eq:yukawa}):
\beq
\Lambda \gaem 11 \, {\rm  TeV}
\sqrt{\kappa\left({m_c\over 1.5\, {\rm GeV}}\right)}~,
\label{eq:Dbound}
\eeq
so that $\Lambda \gaem 21$ TeV for $\kappa \approx 3.5$.

The $\Delta C = 2$, LR product of color-singlet currents gives a weaker
bound than eqn.~(\ref{eq:Dbound}) but the LR product of color-octet currents,
\beq
{\cal L}_{eff} = - \,
\cos \theta_L^c \sin \theta_L^c \cos \theta_R^c \sin \theta_R^c
\frac{g^2}{M^2}
( \overline c_L \gamma^{\mu} T^a u_L)
(\overline c_R \gamma_{\mu} T^a u_R) ~,
\label{ops2}
\eeq
where $T^a$ are the generators of $SU(3)_C$, gives a stronger bound:
\beqa
\left(\frac{M}{g}\right)_{ \! {\rm LR}} & \gaem &
\frac{4 f_D}{3(m_c + m_u)} \left( \frac{m_D^3 B_D^\prime}{\Delta
  m_D}\right)^{\! 1/2}
(\cos \theta_L^c \sin \theta_L^c \cos \theta_R^c \sin \theta_R^c)^{\!
  1/2}
\\ [2mm]
& \approx & 240 \, {\rm TeV} \left({1.5\, {\rm GeV}\over m_c}\right)~,
\eeqa
corresponding to 
\beq
\Lambda \gaem 22 \, {\rm  TeV}
\sqrt{\kappa\left({1.5\, {\rm GeV}\over m_c}\right)}~.
\label{eq:DDbound}
\eeq

\section{Higgs Mass Limits}

Because of triviality, a lower bound on the scale $\Lambda$ yields an
upper limit on the Higgs boson's mass. A rigorous determination of this
limit would require a nonperturbative calculation of the Higgs mass in
an $O(4)$-symmetric theory subject to the constraint on $\Lambda$.  Here
we use eqn.~(\ref{estimate}) to provide an estimate of
this upper limit by naive extrapolation of the lowest-order perturbative
result.\footnote{The naive perturbative bound has been remarkably close
  to the non-perturbative estimates derived from lattice Monte Carlo
  calculations.\cite{lattice}}  The bound $\Lambda \gaem 13$ TeV given
by the contribution of the $\Delta S = 2$ product of color-singlet
currents to the $K_L - K_S$ mass difference, eqn.~(\ref{eq:fcncbound}),
in the case $\kappa \approx 3.5$, results in the limit\,\footnote{If
  $\kappa \approx 4\pi$, $\Lambda$ would have to be greater than 24 TeV,
  yielding an upper limit on the Higgs boson's mass of 450 GeV. If
  $\kappa \approx 1$, $\Lambda$ would be greater than 6.8 TeV, yielding
  the upper limit $m_H \laem 570$ GeV.} $m_H \laem 490$ GeV.  The bound
$\Lambda \gaem 21$ TeV, given by the contribution of the $\Delta C = 2
\,$, LL or RR product of color-singlet currents to the neutral $D$-meson
mass difference, eqn.~(\ref{eq:Dbound}), yields $m_H \laem 460$ GeV.
Limits from the contributions of color-octet currents or from the
relationship between $m_c$ and $\Delta m_K$ would be even more
stringent.

\section{Conclusions}

Because of triviality, theories with a heavy Higgs boson are effective
low-energy theories valid below some cut-off scale $\Lambda$.  We have
shown that the experimental constraint on the amount of custodial
symmetry violation implies that the scale $\Lambda$ must be greater than
of order 7.5 TeV.  The underlying high-energy theory must also include
flavor dynamics at a scale of order $\Lambda$ or greater in order to
produce the different Yukawa couplings of the Higgs to ordinary
fermions.  This flavor dynamics will generically give rise to
flavor-changing neutral currents.  In this note we showed that
satisfying the experimental constraints on extra contributions to
$\Delta m_K$ and $\Delta m_D$ requires that the scale of the associated
flavor dynamics exceed certain lower bounds. At the same time, the new
physics must provide sufficiently large Yukawa couplings to give the
quarks their observed masses.  In order to give rise to a sufficiently
large $s$-quark Yukawa coupling, we showed that $\Lambda$ must be
greater than of order 13 TeV, while in the case of the $c$-quark the
bound is even more stringent, $\Lambda \gaem 21$ TeV.  

For theories defined about the infrared-stable Gaussian fixed-point, we
estimated that this lower bound on $\Lambda$ yields an upper limit of
approximately 460 GeV on the Higgs boson's mass, independent of the
regulator chosen to define the theory. We also showed that some
regulator schemes, such as higher-derivative regulators, used to define
the theory about a different fixed-point are particularly dangerous
because an infinite number of custodial-isospin-violating operators
become relevant.

\vspace{12pt} \centerline{\bf Acknowledgments} \vspace{10pt}

R.S.C. and E.H.S. thank Koichi Yamawaki and the organizers of SCGT 96
for holding a stimulating conference. E.H.S. acknowledges the support of
the NSF Faculty Early Career Development (CAREER) program, the DOE
Outstanding Junior Investigator program, and the JSPS Invitation
Fellowship Program. {\em This work was supported in part by the National
  Science Foundation under grant PHY-9501249, and by the Department of
  Energy under grant DE-FG02-91ER40676.}

\end{document}